# Time-reversal and parity violating superconducting state of $Bi_2Sr_2CaCu_2O_8$

Partha Goswami

*Physics Department, Deshbandhu College, Kalkaji, New Delhi-110019, India.*

*Email of the corresponding author:physicsgoswami @gmail.com*

**Abstract.** The time-reversal and parity violating superconducting(SC) phase of $Bi_2Sr_2CaCu_2O_8$ induced by a magnetic field (B), reported by Krishana et al[K. Krishana et al, Science 277, 83 (1997)] over the limited field range of 0.6T < B < 5T, is examined starting with the model free-energy functional predicted by Laughlin [R. B. Laughlin, Physica C 234, 280 (1994)]. On the basis of entropy considerations we show that the passage from the normal to this field-driven SC state at a non-zero temperatures (T) is possible if $k_BT \leq \Delta_0$ where the zero-temperature gap $\Delta_0 \sim \hbar v/l_B$ , v is the root-mean-square velocity of the d-wave node, and $l_B = \sqrt{(\hbar/eB)}$ is the magnetic length. The restriction may be construed as setting lower limit on the required field strength at a given temperature. We find that, over the field range mentioned, the specific heat coefficient exhibits $\sqrt{B}$- and the B-linear dependence (Volovick effect).

## INTRODUCTION

Nearly a decade and a half ago Krishna et al.[1] have reported a superconducting(SC) phase transition in $Bi_2Sr_2CaCu_2O_8$ induced by a magnetic field(B) over the limited field range of 0.6T < B < 5T. The transition was observed to be characterized by a kink in the thermal conductivity as a function of field strength followed by a plateau in SC state. They argued that the plateau signifies a halt in the heat transport by quasi-particles in the SC state due to the appearance of a gap $\Delta$ in the excitation spectrum. The authors also reported the empirical relation $T_c$ proportional to $\sqrt{B}$. Laughlin[2] had suggested that in this class of superconductors the origin of the induced energy gap $\Delta$ lies in the development of a small $d_{xy}$ superconducting order parameter phased by $\pi$ /2 with respect to the principal $d_{x2\,-y2}$ one leading to the violation of both parity and time-reversal symmetry. The violation makes these superconductors carry gap induced magnetic moment and quantized hall conductance at near zero temperature. He hypothesized a model free-energy functional[2] which is the starting point of this communication. Including an additional correction term of $O(\Delta^5)$ in the functional, we find that the passage from the normal to the superconducting state at non-zero temperatures (T) is possible if $k_BT \leq \Delta_0$,where $\Delta_0 = 1.329\ (\hbar v/l_B)$. We find that the entropy is vanishingly small till $\Delta/\Delta_0 \sim 2$ and positive beyond this value as long as $k_BT \leq \Delta_0$. For $k_BT > \Delta_0$ we obtain unphysical negative entropy beyond $\Delta/\Delta_0 \sim 2$ (see Figure 1(a)). Taking $(\hbar v) = 0.30$ eV A$^o$[3] we find that the restriction implies B $\geq$ 0.52 Tesla for T = 10 K. The restriction $k_BT \leq \Delta_0$ may be interpreted as setting lower limit on the required field strength at a given temperature. We also find that, over the range 0.5T < B < 2T, the specific heat coefficient exhibits Volovick effect [4] ($\sqrt{B}$-and the B-linear dependence). (see Figure 1(b)).

## FREE ENERGY FUNCTIONAL

We consider the Laughlin free-energy functional [2] with the inclusion of a term of $O(\Delta^5)$. The functional is

$$(0.139F/|F_0|)=1.060\ (\Delta/\Delta_0)^5 + (1/2)\ (\Delta/\Delta_0)^3 - 1.699(\Delta/\Delta_0)\tanh^2(\beta\Delta/2)+11.978\ (k_B T/\Delta_0)^3 \times\{-(\beta\Delta)^2/2\times\ln(1+\exp(-\beta\Delta))+\int_0^{\beta\Delta} \ln(1+\exp(-x))\ x\ dx\} \quad (1)$$

where the zero-temperature free energy is $(|F_0|/L^2)=((\hbar v)^{-2}\Delta_0^3/21.5517\pi)$. The dimensionless entropy is given by $S = \beta\,(\partial(F/|F_0|)/\partial\beta)$. We obtain

$$S \approx -(3/2)((\beta\Delta)^2/(\beta\Delta_0))\tanh(\beta\Delta/2) \times \mathrm{sech}^2(\beta\Delta/2) + 36(k_B T/\Delta_0)^3 \times \{ (\beta\Delta)^2/6 \times \ln(1+\exp(-\beta\Delta)) + (\beta\Delta)^3/6 \times (1+\exp(\beta\Delta))^{-1} - 0.116(\beta\Delta)^2 + (\beta\Delta)^4/96 \}. \quad (2)$$

The specific heat, on the other hand, may be written as $C \approx \gamma(B,T)\,T$ where the coefficient

$$\gamma(B,T) = (3k_B/2)((\beta\Delta)^2/\Delta_0)\tanh(\beta\Delta/2) \times \mathrm{sech}^2(\beta\Delta/2) + (3k_B/4)((\beta\Delta)^3/\Delta_0) \times \mathrm{sech}^4(\beta\Delta/2) + (3k_B/2)((\beta\Delta)^3/\Delta_0) \times \tanh^2(\beta\Delta/2)\,\mathrm{sech}^2(\beta\Delta/2) + 108\,\beta k_B(\beta\Delta_0)^{-3} \times \{ (\beta\Delta)^2/6 \times \ln(1+\exp(-\beta\Delta)) + (\beta\Delta)^3/6 \times (1+\exp(\beta\Delta))^{-1} - 0.116(\beta\Delta)^2 + (\beta\Delta)^4/96 \} - 36\beta k_B(\beta\Delta_0)^{-3} \times \{ (\beta\Delta)^2/3 \times \ln(1+\exp(-\beta\Delta)) + (\beta\Delta)^3/3 \times (1+\exp(\beta\Delta))^{-1} - 0.232(\beta\Delta)^2 + (\beta\Delta)^4/32 - (\beta\Delta)^4/6 \times \exp(\beta\Delta) \times (1+\exp(\beta\Delta))^{-2} \}. \quad (3)$$

In Figure 1(a) we have plotted the entropy given by (2) as a function of $(\Delta/\Delta_0)$ for $k_B T/\Delta_0$ = 0.5 (Curve I), 1.0 (Curve II), 2.0 (Curve III),and higher values. We find $S \geq 0$ for $k_B T \leq \Delta_0$ while $S \leq 0$ for $k_B T > \Delta_0$ as already stated. In Figure 1(b) we have plotted the specific heat coefficient given by (3) versus $B^{1/2}$ at T= 13.5 K. We find that, over the field-range mentioned above, $\gamma(B)$ exhibits Volovick effect [4]($\sqrt{B}$- and the B-linear dependence). This is, as expected, in contrast to the usual $\sqrt{B}$-dependence of a clean d-wave superconductor and the B-linear dependence in the ±s-wave state of the iron-based superconductors. The result bears the signature of the gapped nodal regions on the Fermi surface of $Bi_2Sr_2CaCu_2O_8$.

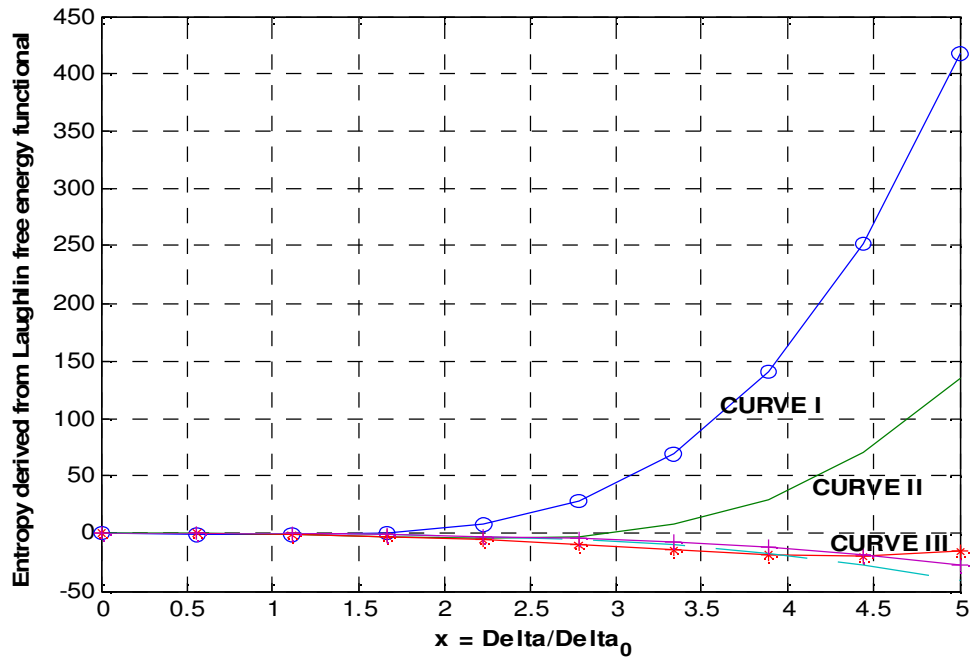

(a)

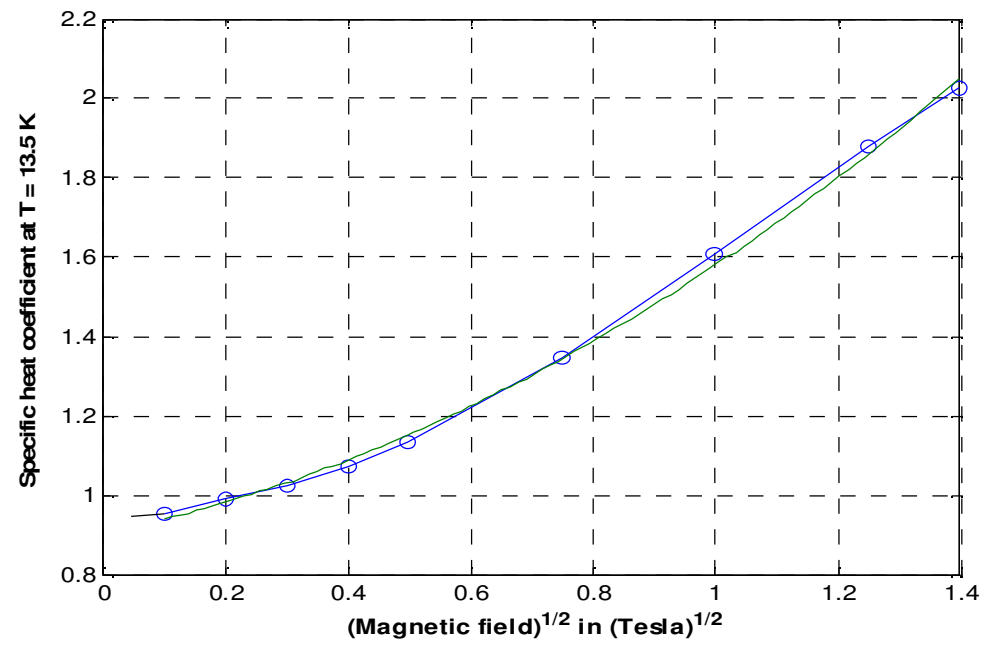

(b)

**FIGURE 1.** (a) A plot of the entropy derived from the free energy functional (1) as a function of $\Delta/\Delta_0$ for $k_B T/\Delta_0$ = 0.5,1.0, 5.0, 10, and 15. (b) A plot of the specific heat coefficient versus $B^{1/2}$ at T =13.5 K and a second degree polynomial fit.